\documentclass[journal]{IEEEtran}

\ifCLASSINFOpdf
\else
   \usepackage[dvips]{graphicx}
\fi
\usepackage{url}

\usepackage{graphicx}
\usepackage{cite}
\usepackage{multirow}
\usepackage{color}
\usepackage{balance}
\usepackage{xcolor}

\begin{document}
	\title{Fast, nonlocal and neural: a lightweight high quality solution to image denoising} 

	\author{{Yu~Guo,~\IEEEmembership{Graduate Student Member,~IEEE,} Axel~Davy, Gabriele~Facciolo, \\ Jean-Michel~Morel and Qiyu~Jin,~\IEEEmembership{Member,~IEEE}}
		\thanks{This work was supported by National Natural Science Foundation of China (No. 12061052), Natural Science Fund of Inner Mongolia Autonomous Region (No. 2020MS01002), Prof. Guoqing Chen's “111 project”of higher education talent training in Inner Mongolia Autonomous Region and the network information center of Inner Mongolia University. Work partly financed by Office of Naval research  grants N00014-17-1-2552 and N00014-20-S-B001,  DGA Defals challenge n$^\circ$ ANR-16-DEFA-0004-01, MENRT  and Fondation Mathématique Jacques Hadamard. \textit{(Corresponding author: Qiyu~Jin.)}}
		\thanks{Y. Guo and Q. Jin are with School of Mathematical Science, Inner Mongolia University, Hohhot 010020, China (e-mail: yuguomath@aliyun.com; qyjin2015@aliyun.com).}
		\thanks{A. Davy, G. Facciolo and J-M. Morel are with Centre Borelli, ENS Paris-Saclay, CNRS, 4, avenue des Sciences 91190 Gif-sur-Yvette, France (e-mail: axel.davy@normalesup.org; gabriele.facciolo@ens-paris-saclay.fr; morel@ens-paris-saclay.fr).}}

	\markboth{Manuscript to IEEE SIGNAL PROCESSING LETTERS 2021}
	{Shell \MakeLowercase{\textit{et al.}}: Bare Demo of IEEEtran.cls for IEEE Journals}
	\maketitle
	\begin{abstract}
		With the widespread application of convolutional neural networks (CNNs), the traditional model based denoising algorithms are now outperformed. However, CNNs face two problems. First, they are computationally demanding, which makes their deployment especially difficult for mobile terminals. Second, experimental evidence shows that CNNs often over-smooth regular textures present in images, in contrast to traditional non-local models. In this letter, we propose a solution to both issues by combining a nonlocal algorithm with a lightweight residual CNN. This solution gives full latitude to the advantages of both models. We apply this framework to two GPU implementations of classic nonlocal algorithms (NLM and BM3D) and observe a substantial gain in both cases, performing better than the state-of-the-art with low computational requirements. Our solution is between 10 and 20 times faster than CNNs with equivalent performance and attains higher PSNR. In addition the final method shows a notable gain on images containing complex textures like the ones of the \emph{MIT Moir\'{e}} dataset. 
	    
	\end{abstract}


	\IEEEpeerreviewmaketitle

	\section{Introduction}
	Denoising is one of the most critical issues in image processing.  Indeed, noise  lowers the final image quality and impacts all downstream computer vision tasks. After a variance stabilizing transform (VST) the raw observed image can be expressed as 
	\begin{equation}
	y = x + n,
	\end{equation}
	where $y$ is the observed image, $x$ is the underlying clean image, and $n$ is an additive approximately white Gaussian noise (AWGN). In the past few decades and before the emergence of CNNs, several image prior models have been proposed, particularly the  non-local models (NLM~\cite{buades2005review}, NLBayes~\cite{Lebrun2013NLBayes}, BM3D~\cite{Dabov2007BM3D}, WNNM~\cite{Gu2014WNNM}) and sparse models (LSSC~\cite{Mairal2009LSSC}). These algorithms restore well textures and repeated structures, but they also leave behind residual noise and  cause blur.
	
	The non-local models are now clearly outperformed by data-driven CNNs such as DnCNN~\cite{Zhang2017DnCNN}, FFDNet~\cite{Zhang2018FFDNet} and others \cite{TIAN2020ADNet,Jang2020Dual,Song2020Grouped,Wang2021Channel}. Yet
	these data-driven algorithms are energy greedy and require high-performance GPUs.
	This hinders their deployment on a large scale, especially for mobile terminals such as mobile phones. Furthermore, although CNNs show excellent PNSR performance, it has been observed that they over-smooth regular textures and self-similar structure~\cite{Cruz2018NN3D}. 

Aware of the scale and speed of CNNs, Gu et al.~\cite{Gu2019self} proposed a self-guided fast network using multi-resolution input. 
Ma et al.~\cite{Ma2020Pyramid} combined a pyramid neural network with the Two-Pathway Unscented Kalman Filter and proposed a fast algorithm for real-world noise removal. 
Aiming at raw image denoising of mobile devices, Wang et al.~\cite{Wang2020Practical} proposed a lightweight and effective neural network and k-Sigma noise transformation method.

The second problem has led many researchers to incorporate non-local routines into their CNN design. 
Wang \cite{wang2016small} proposed a non-convolutional denoising lightweight network trained on texture patches. 
The texture patches are detected,  denoised by the network and then aggregated to the BM3D result. 
Yang et al.~\cite{Yang2018BM3D-Net} referred to the BM3D algorithm and proposed   ``extraction" and ``aggregation" layers to model the block matching in BM3D, and used CNNs to replace the collaborative filtering stage.
Cruz et al.~\cite{Cruz2018NN3D} placed plug-and-play non-local filters after CNNs, thus adding a nonlocal regularity term.  
In \cite{Liu2018NLRNet} the authors, inspired by non-local neural networks \cite{Wang2018Non-local}, introduced non-local CNNs into image restoration. 
Lefkimmiatis \cite{Lefkimmiatis2017NCDNet} performed block matching and weighted non-local sum on the results of 2D Convolution, thereby effectively integrating non-locality into CNNs.
Taking advantage of the relaxation of the k-nearest neighbors matching selection rule, Pl\"{o}tz and Roth~\cite{2018N3Net} proposed a novel non-local layer.  
All of the above mentioned papers gave evidence that a non-local routine indeed improves the receptive field of CNNs and enhances their performance. 
Unfortunately, given the high dimensionality of their feature space, non-local operations on CNNs further increase their power consumption. 
	
Both nonlocal algorithms and deep learning based ones have their own advantages and disadvantages, and they complement each other's shortcomings. In this letter, a network is proposed to combine a nonlocal algorithm and a deep learning based method to benefit from  the advantages of both techniques. In the proposed network, the output of a nonlocal method is used as a pre-processing algorithm for it  protects the image details while reducing noise, then a  network is used to learn the residual of the restored image of the nonlocal method. Here, the network corrects the restored image given by the nonlocal method, that is, removes artifacts, color distortions and residual noise from the image, rather than full denoising. As a result, a very lightweight network can produce excellent results.

The rest of the paper is organized as follows. In Section~\ref{II}, we verify the effectiveness of the proposed scheme. We find that preprocessing by BM3D  leads to a drastic reduction of the size of the CNN yielding an optimal solution. These experiments made with fixed noise CNNs lead us to propose  a  lightweight and flexible color image denoising method in Section~\ref{III}.  This algorithm  obtains high-quality results at a lower computational cost for images with variable noise. Experiments evaluating image quality and computational cost are presented in Section~\ref{IV}, where our algorithm is compared with state-of-the-art denoising algorithms.

	\begin{table}[t]
		\caption{
		CPSNR(dB) comparison between our proposed method and the state-of-the-art methods with the fixed parameter.
		}
		\label{Tab1}
		\begin{center}
	    \addtolength{\tabcolsep}{-4pt}
			\begin{tabular}{ccccccccc}
				\hline
    			\multirow{2}{*}{Dataset}  & \multirow{2}{*}{$\sigma$} & BM3D & BM3D & DnCNN	& ADNet  & DnCNN & Ours & Ours  \\ 
    			 &  & \cite{Dabov2007CBM3D} & -G~\cite{Davy2021BM3D} & \cite{Zhang2017DnCNN}	& \cite{TIAN2020ADNet}  & $K = 10$ & $K = 10$ & $K = 16$ \\ \hline
    			 
                & 25  & 31.88 & 31.72  & 32.34 & 32.30 & 32.08 & 32.38  & \textbf{32.40}\\
	            \textit{Kodak} & 35 & 30.29 & 30.14 & 30.75 & 30.75 & 30.47 & 30.82 & \textbf{30.86} \\
				\cite{zhang2011color}  & 50 & 28.68 & 28.55 & 29.15 & 29.18 & 28.85 & 29.26 & \textbf{29.32} \\
				& 75 & 26.98 & 26.88 & --    & 27.51 & 27.08 & 27.56 & \textbf{27.64} \\ \hline
				
				& 25 & 31.72 & 31.65 & 32.49 & 32.47 & 32.23 & 32.52 & \textbf{32.55} \\
				\textit{McM} & 35 & 30.18 & 30.11 & 30.93 & 30.93 & 30.60 & 30.96 & \textbf{31.02}  \\
				\cite{Dubois2005Frequency} & 50 & 28.53 & 28.48 & 29.23 & 29.29 & 28.92 & 29.33 & \textbf{29.43} \\
				& 75 & 26.76 & 26.72 & --    & 27.47 & 27.06 & 27.48 & \textbf{27.59} \\ \hline
				
				& 25 & 30.76 & 30.62 & \textbf{31.25} & 31.19 & 31.06 & 31.23 & \textbf{31.25} \\
				\textit{CBSD68} & 35 & 29.09 & 28.94 & 29.59 & 29.56 & 29.37 & 29.59 & \textbf{29.62} \\
				\cite{2009Fields} & 50 & 27.48 & 27.35 & 28.01 & 27.95 & 27.73 & 27.97 & \textbf{28.02} \\
				& 75 & 25.69 & 25.60 &--     & 26.24 & 25.95 & 26.23 & \textbf{26.30} \\ \hline
				
				& 25 & 30.41 & 30.39 & 29.92 & 29.97 & 29.63 & 30.59 & \textbf{30.64} \\
				\textit{MIT} & 35 & 28.82 & 28.80 & 28.34 & 28.41 & 28.04 & 28.99 & \textbf{29.07} \\
				\cite{gharbi2016deep} & 50 & 27.07 & 27.06 & 26.76 & 26.86 & 26.47 & 27.38 & \textbf{27.46} \\
				& 75 & 25.34 & 25.36 & --    & 25.22 & 24.80 & 25.61 & \textbf{25.69} \\ \hline
				
			    & 25 & 30.97 & 30.75 & 31.17 & 31.23 & 30.86 & 31.64 & \textbf{31.68} \\
				\textit{Urban} & 35 & 29.25 & 29.05 & 29.37 & 29.42 & 28.00 & 29.89 & \textbf{29.96} \\
				\cite{Huang2015Single} & 50 & 27.43 & 27.28 & 27.48 & 27.57 & 27.12 & 28.09 & \textbf{28.17} \\
				& 75 & 25.32 & 25.26 & --    & 25.50 & 25.02 & 25.98 & \textbf{26.07} \\ \hline

			\end{tabular}
		\end{center}
	\end{table}

	\section{Combination of BM3D and CNN}
	\label{II}
\begin{figure}[t]
		\centering
		\includegraphics[width=0.42\textwidth]{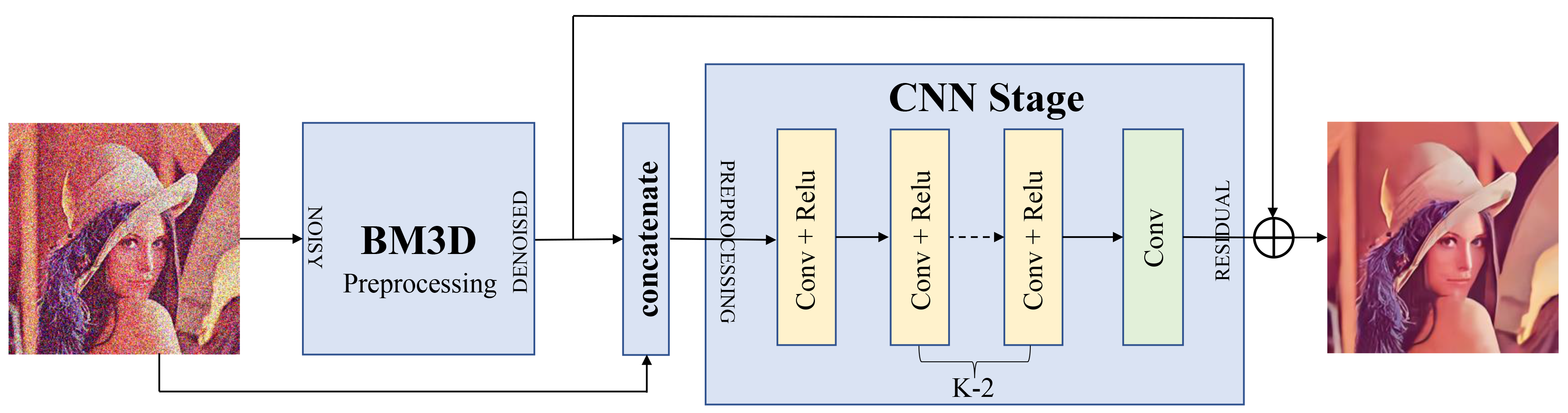}
		\caption{Algorithm structure. Using residual learning to combine BM3D and CNNs.}  
		\label{Algorithm}
	\end{figure}

	We take inspiration in residual learning, which has been used in other areas of image restoration such as demosaicing~\cite{tan2017CDM} and super-resolution~\cite{Dong2014SRCNN}. 
	We test a processing pipeline where a fast, nonlocal and neural filter first removes noise by a GPU-accelerated BM3D algorithm (BM3D-G)~\cite{Davy2021BM3D}. Then we feed both the restored image and noisy image to the CNN to learn a residual, finally obtaining the restored image by adding the residual to the BM3D  preprocessed image.  This specific model is shown in Fig. \ref{Algorithm}.

	Using the BM3D-G preprocessing step reduces floating point operations and lightens the network.
	With reference to the simple and effective design of DnCNN~\cite{Zhang2017DnCNN}, $3\times3$ convolutions are used to build the network. To test how far BM3D could allow for a reduction of the scale of the network, we test a network with 10 convolutional layers ($K=10$, $44.6\%$ of the DnCNN scale) and a network with 16 convolutional layers ($K=16$, $77.1\%$ of the DnCNN scale).

    The training hyperparameters for our network are the same as \cite{Zhang2017DnCNN}. The training dataset is the Waterloo Exploration Database (WED) \cite{Ma2017Waterloo}, which contains $4744$ natural images. We crop the images into $200,000$ $128\times128$ patches and used an $L_1$ loss defined by
	\begin{equation}
	\mathcal{L}(\Theta)=\frac{1}{N}\sum_{i=1}^N|x_i-[F(z_i;\Theta)+BM3D(y_i)]|,
	\end{equation}
	\begin{equation}
	z_i=Concatenate(BM3D(y_i),y_i),
	\end{equation}
	where $y_i$ is a noisy image and $x_i$ is the corresponding clean image. 
	We considered five datasets to comprehensively test the performance of the algorithm. There are three natural image sets (Kodak~\cite{zhang2011color}, MacMaster~\cite{Dubois2005Frequency}, CBSD68~\cite{2009Fields}) and two texture image sets (MIT moir\'{e}~\cite{gharbi2016deep}, Urban100~\cite{Huang2015Single}). Our comparison criterion is the composite PSNR (CPSNR)~\cite{Alleysson2005CPSNR}.

	In Table \ref{Tab1}, we can see that with only  $K = 10$ convolutional layers, our proposed method  reaches the performance of DnCNN while having half of its size. With $K = 16$ convolutional layers, our proposed scheme outperforms DnCNN. On the texture image sets, our method shows a significant improvement. On the MIT moir\'{e} dataset, the average gain of our proposed method is as high as $0.60$dB, and on Urban100, the CPSNR improves by nearly $0.54$dB. 
	
	\begin{figure}[t]
		\centering
		\renewcommand{\arraystretch}{0.4} \addtolength{\tabcolsep}{-5pt} {%
			\fontsize{7pt}{\baselineskip}\selectfont
			\begin{tabular}{cc}
				\includegraphics[width=0.20\textwidth]{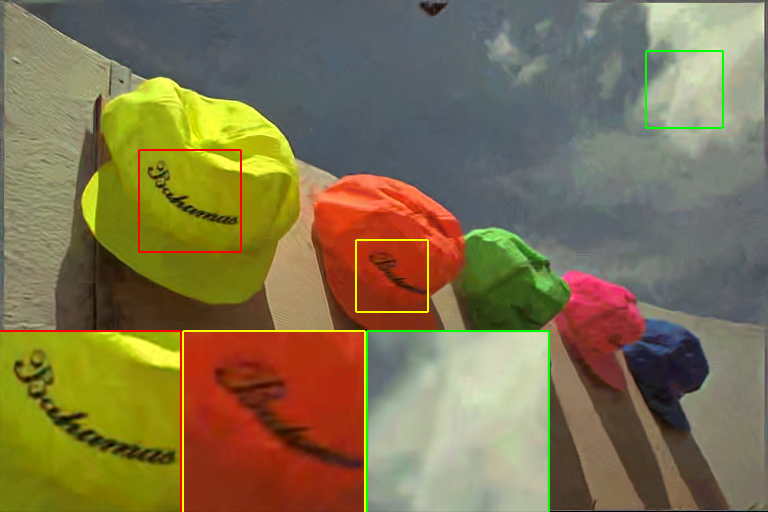} &
				\includegraphics[width=0.20\textwidth]{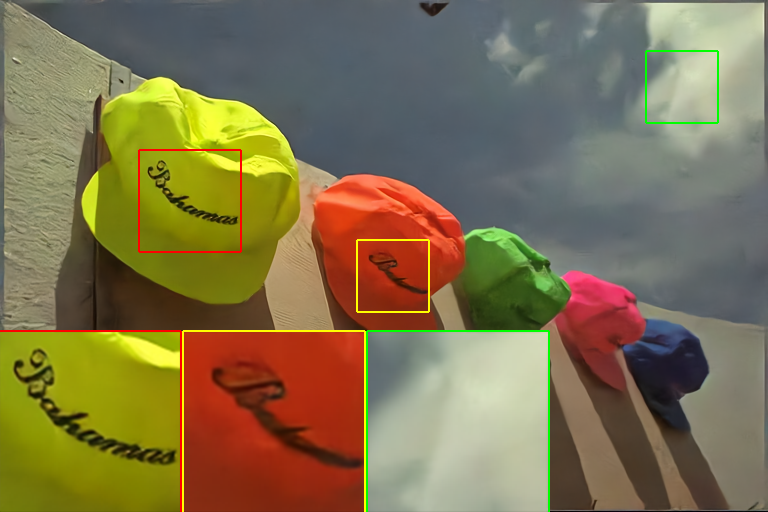} \\
				\includegraphics[width=0.20\textwidth]{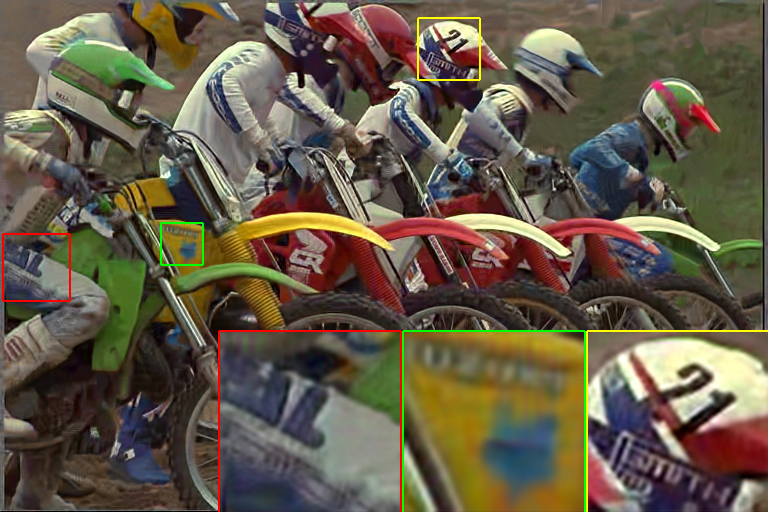} &
				\includegraphics[width=0.20\textwidth]{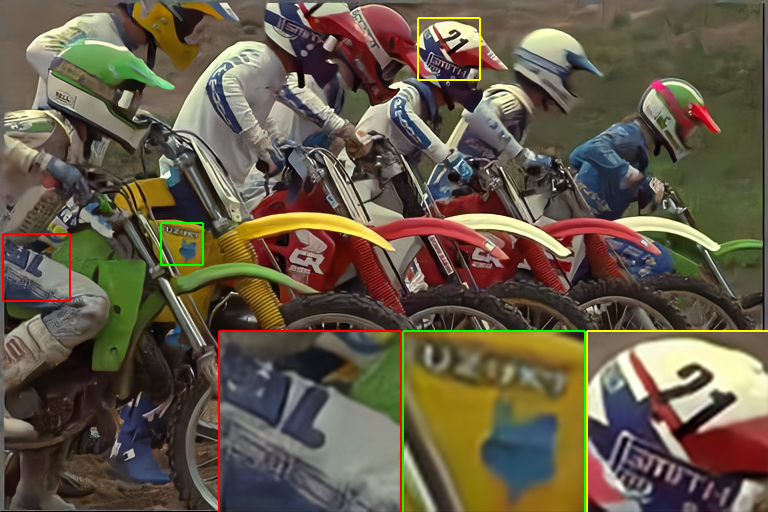} \\
				(a) BM3D-G & (b) Ours ($K = 10$)
			\end{tabular}
		}
		
		\caption{Result of BM3D before and after being post-processed by a lightweight CNN.}
		\label{bm3d}
	\end{figure}
	

	These experiments show that using BM3D to preprocess noisy images has two advantages: 
	\begin{itemize}
		\item After the BM3D preprocessing, the difficulty of solving the inverse problem is reduced. This explains why the network can be shallower without performance loss. 
		
		\item 
	    The use of BM3D preprocessing improves the non-local performance of the network, with a just small increase in cost. The combination greatly improves network effectiveness in processing texture images or  self-similar structures. 
	\end{itemize}
	
	Although BM3D has remarkable performance, it introduces  unpleasant artifacts, that become conspicuous at higher noise levels. Fig.~\ref{bm3d} (a)  illustrates that there are ripples, color distortion and blur in the BM3D restored images. The CNN effectively corrects these artifacts and provides pleasant restored images (see Fig.~\ref{bm3d} (b)).

	In summary, the combination of BM3D and CNN not only overcomes each other's shortcomings  but also improves the quality of denoised images. Yet these experiments were conducted with a fixed noise model, requiring  training for each noise level. This is not practical for mobile devices. In the next section we address this problem.

	\begin{figure}[t]
	\centering
	\includegraphics[width=0.48\textwidth]{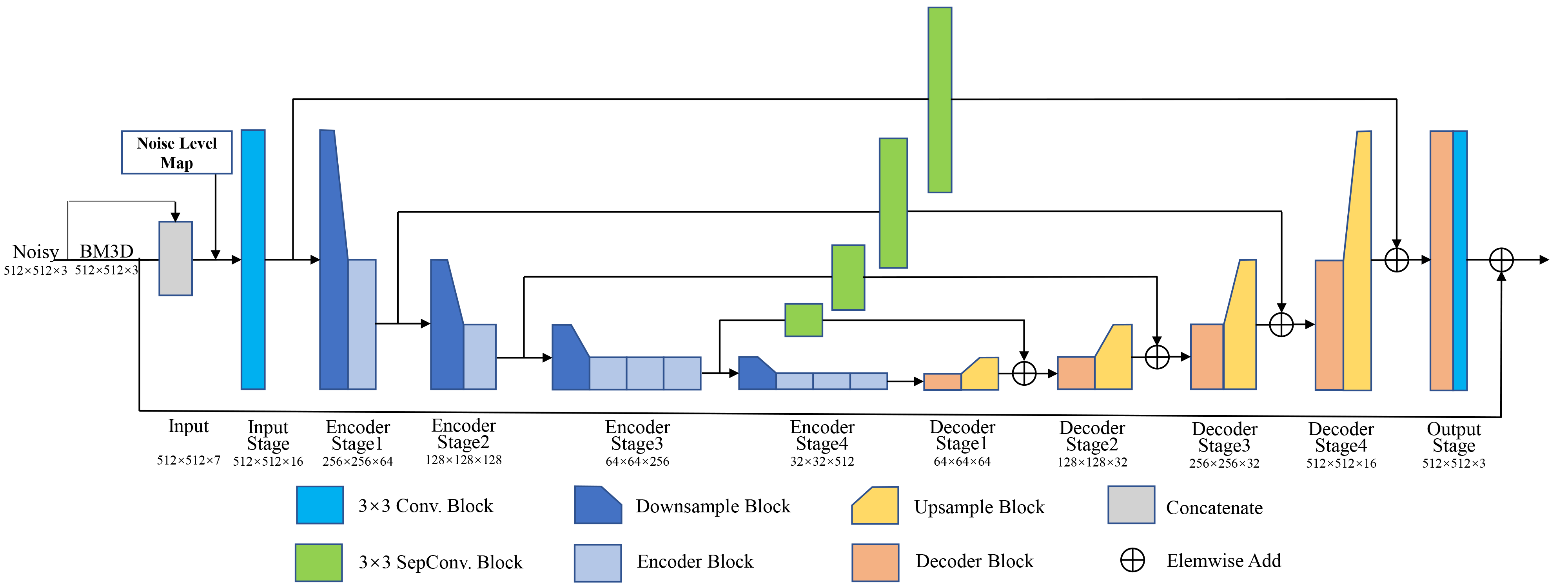} \\
	\includegraphics[width=0.48\textwidth]{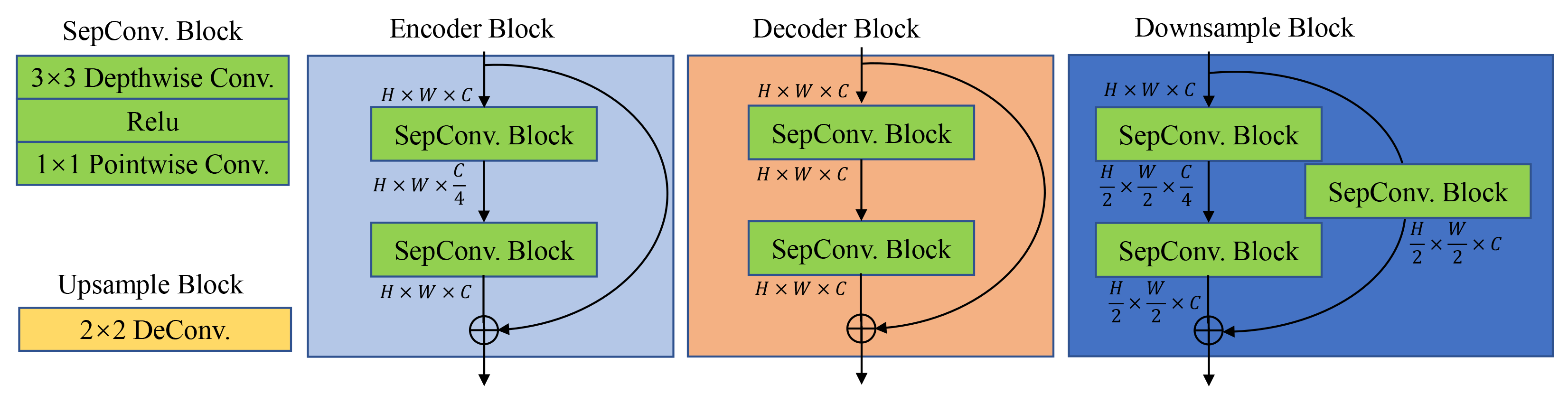}
	\caption{Flexible and lightweight U-Net-like  structure of denoising.} 
	\label{U}
	\end{figure}

	\section{A flexible and Lightweight Denoising Algorithm}
	\label{III}

     CNNs trained for a fixed noise level perform well but cannot be practically applied in real scenes due to huge memory requirements. 
     In order to be widely used in mobile terminals (such as mobile phones), an algorithm has to satisfy two conditions:
	\begin{itemize}
		\item Flexibility: The algorithm must be able to deal with various levels of noise flexibly, instead of training for each level of noise like most CNNs, as done in~\cite{Zhang2018FFDNet}.
		\item Lightweight: Most devices do not have high-performance GPUs. So the algorithm must be lightweight  to ensure high-speed operation in a low-power environment.
	\end{itemize}

	Considering the above two characteristics and the conclusion of Sec. \ref{II}, we designed a flexible and lightweight U-Net-like denoising algorithm.
	The overall network structure is shown in Fig.~\ref{U}. We refer to the design of FFDNet~\cite{Zhang2018FFDNet} to introduce a noise level map to deal flexibly with noise. We use a U-Net architecture~\cite{Ronneberger2015UNet} with 4 encoders and 4 decoders. In order to preserve the low-dimensional characteristics of the network, skip connections are added. The entire network uses $3\times3$ depthwise separable convolutions~\cite{Andrew2017MobileNets} to reduce computational consumption. The stride-2 depthwise separate convolution is used for downsampling, and the corresponding upsampling uses a $2\times2$ deconvolution. The details of the network block are shown in Fig.~\ref{U}. See Sec.~\ref{II} for the relevant settings of network training and testing. 
	
	\begin{table}[t]
		\begin{center}
		\caption{
		CPSNR(dB) value Comparison between our proposed method and the state-of-the-art methods with the flexible parameter.}
		\label{Tab2}
		  \renewcommand{\arraystretch}{1} \addtolength{\tabcolsep}{-4.1pt}
			\begin{tabular}{ccccccccc}
				\hline 
				\multirow{2}{*}{Dataset}  & \multirow{2}{*}{$\sigma$} & NLM & BM3D & FDnCNN & FFDNet & PMRID & Ours-FL & Ours-FL \\
				
				& & -G\cite{Davy2021BM3D} & -G\cite{Davy2021BM3D} & \cite{Zhang2017DnCNN} & \cite{Zhang2018FFDNet} & \cite{Wang2020Practical} & NLM & BM3D  \\ \hline
				
				& 25 & 30.07 & 31.72 & 32.24 & 32.25 & 32.01 & 32.11 & \textbf{32.26}  \\
				\textit{Kodak} & 35 & 28.72 & 30.14 & 30.68 & 30.69 & 30.53 & 30.65 & \textbf{30.77}  \\
				\cite{zhang2011color} & 50 & 27.20 & 28.55 & 29.07 & 29.10 & 29.04 & 29.16 & \textbf{29.25}  \\
				& 75 & 25.47 & 26.88 & 27.39 & 27.43 & 27.48 & 27.56 & \textbf{27.65}  \\\hline

				& 25 & 29.79 & 31.65 & \textbf{32.39} & 32.36 & 31.74 & 31.91 & 32.22 \\
				\textit{McM} & 35 & 28.42 & 30.11 & \textbf{30.84} & 30.83 & 30.36 & 30.53 & 30.79  \\
				\cite{Dubois2005Frequency} & 50 & 26.86 & 28.48 & 29.18          & 27.29 & 28.90 & 29.05 & \textbf{29.24}  \\
				& 75 & 25.04 & 26.72 & 27.36          & 27.37 & 27.28 & 27.38 & \textbf{27.50}  \\\hline
				
				& 25 & 28.83 & 30.62 & 31.13 & \textbf{31.14} & 30.98 & 31.01 & 31.13 \\
				\textit{CBSD68} & 35 & 27.56 & 28.94 & 29.51 & 29.51          & 29.40 & 29.45 & \textbf{29.54} \\
				\cite{2009Fields} & 50 & 26.06 & 27.35 & 27.95 & 27.95          & 27.85 & 27.91 & \textbf{27.98} \\
				& 75 & 24.36 & 25.59 & 26.16 & 26.17          & 26.21 & 26.26 & \textbf{26.32}  \\\hline
				
				& 25 & 28.49 & 30.39 & 29.74 & 29.73 & 29.16 & 29.91 & \textbf{30.53} \\
				\textit{MIT} & 35 & 27.16 & 28.80 & 28.21 & 28.23 & 27.80 & 28.44 & \textbf{29.02}  \\
				\cite{gharbi2016deep} & 50 & 25.58 & 27.06 & 26.69 & 26.72 & 26.44 & 26.94 & \textbf{27.46}  \\
				& 75 & 23.73 & 25.36 & 25.04 & 25.08 & 25.01 & 25.30 & \textbf{25.76}  \\\hline
				
				& 25 & 28.75 & 30.75 & 30.99 & 30.87 & 29.94 & 30.71 & \textbf{31.22} \\
				\textit{Urban} & 35 & 27.41 & 29.05 & 29.22 & 29.14 & 28.31 & 29.12 & \textbf{29.60}  \\
				\cite{Huang2015Single} & 50 & 25.71 & 27.28 & 27.41 & 27.36 & 26.64 & 27.37 & \textbf{27.88}  \\
				& 75 & 23.57 & 25.26 & 25.27 & 25.26 & 24.80 & 25.35 & \textbf{25.89}  \\\hline
				
				& 25 & 36.09 & 37.70 & 38.15 & 38.34 & 38.51 & 38.55 & \textbf{38.66} \\
				\textit{SIDD} & 35 & 34.49 & 36.20 & 36.76 & 36.99 & 37.35 & 37.37 & \textbf{37.48}  \\
				\cite{Abdelhamed2018} & 50 & 32.65 & 34.44 & 35.19 & 35.44 & 36.04 & 36.03 & \textbf{36.11}  \\
				& 75 & 30.40 & 32.30 & 33.30 & 33.63 & 34.41 & 34.40 & \textbf{34.43}  \\\hline
			
			\end{tabular}
		\end{center}
	\end{table}
	
	\section{Experiments and Analysis}
	\label{IV}
	

	\begin{figure*}[ht]
	\centering
	\renewcommand{\arraystretch}{0.5} \addtolength{\tabcolsep}{-4pt} {%
		\fontsize{7pt}{\baselineskip}\selectfont
		\begin{tabular}{ccccccccc}
		\multirow{1}{*}[26pt]{(A)} &
			\includegraphics[width=0.11\textwidth]{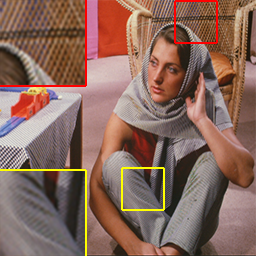} &
			\includegraphics[width=0.11\textwidth]{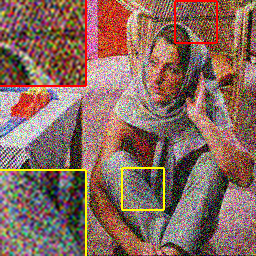} &
			\includegraphics[width=0.11\textwidth]{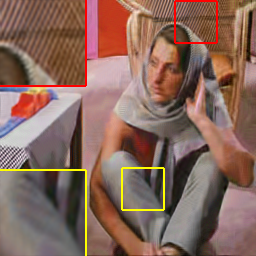} &
			\includegraphics[width=0.11\textwidth]{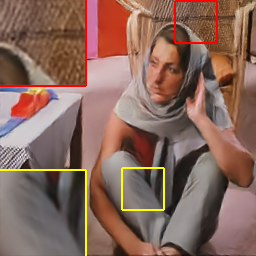} &
			\includegraphics[width=0.11\textwidth]{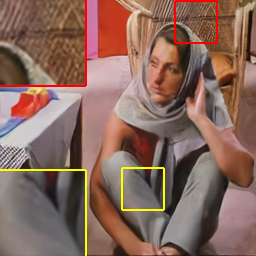} &
			\includegraphics[width=0.11\textwidth]{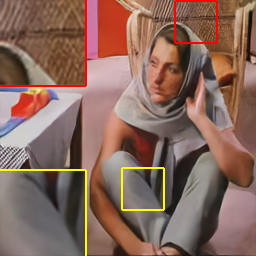} &
			\includegraphics[width=0.11\textwidth]{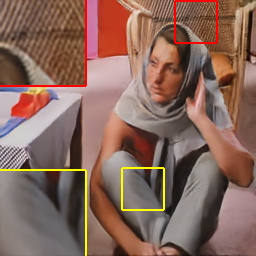} &
			\includegraphics[width=0.11\textwidth]{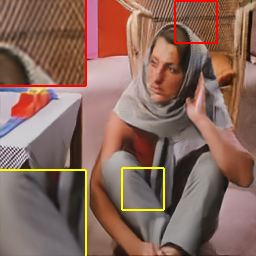}\\
		& (a) Ground truth & (b) Noise ($\sigma=50$) & (c) BM3D-G~\cite{Davy2021BM3D} & (d) PMRID~\cite{Wang2020Practical} & (e) FDnCNN~\cite{Zhang2017DnCNN} & (f) FFDNet~\cite{Zhang2018FFDNet} & (g) Ours-FL (NLM) & (h) Ours-FL (BM3D) \\
		& CPSNR & 14.78 dB & 26.81 dB & 26.82 dB & 27.27 dB & 27.26 dB & 27.27 dB & \textbf{27.36} dB \\

        \multirow{1}{*}[20pt]{(B)} &
			\includegraphics[width=0.11\textwidth]{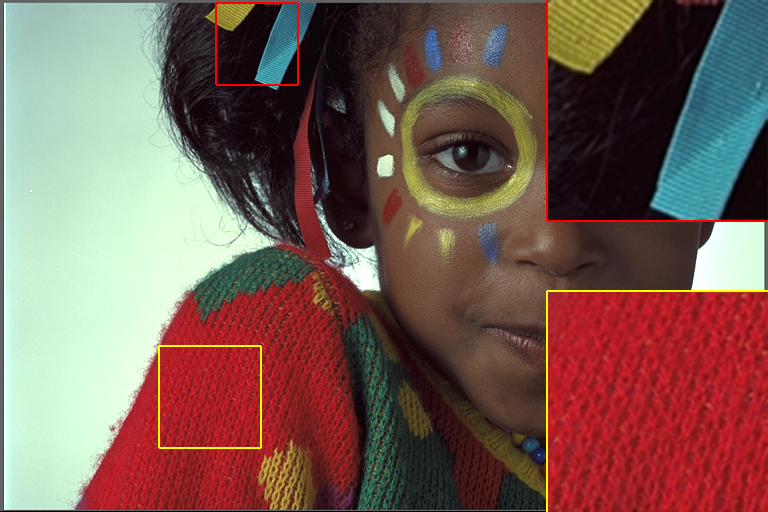} &
			\includegraphics[width=0.11\textwidth]{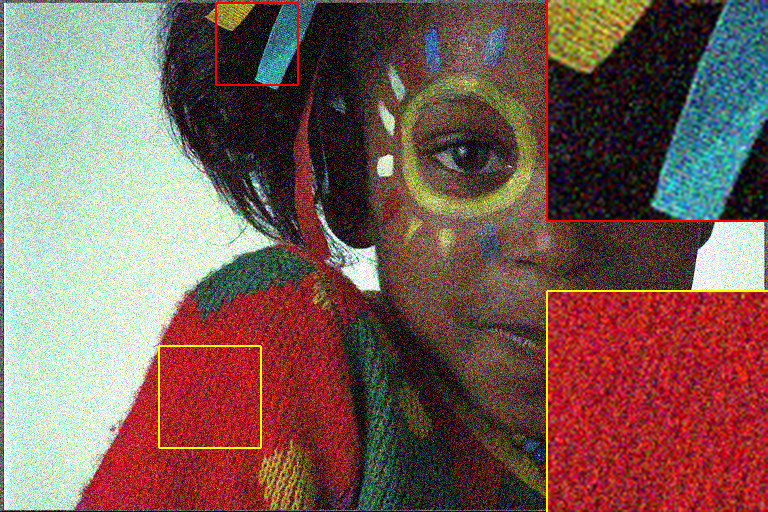} &
			\includegraphics[width=0.11\textwidth]{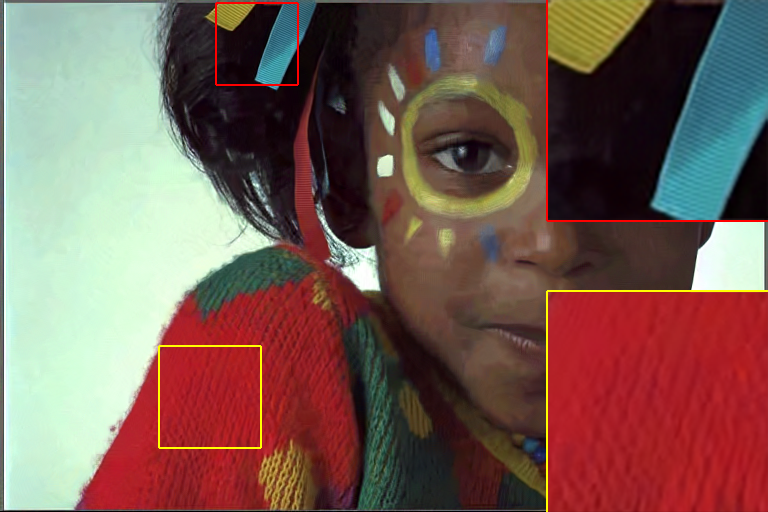}&
			\includegraphics[width=0.11\textwidth]{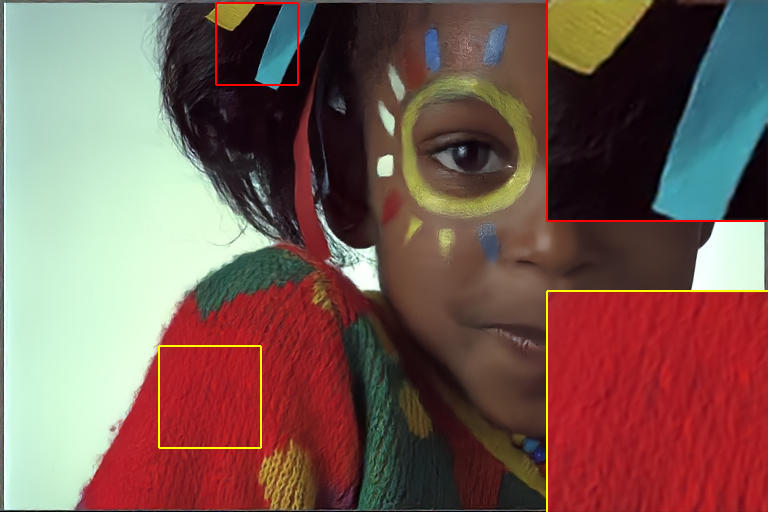} &
			\includegraphics[width=0.11\textwidth]{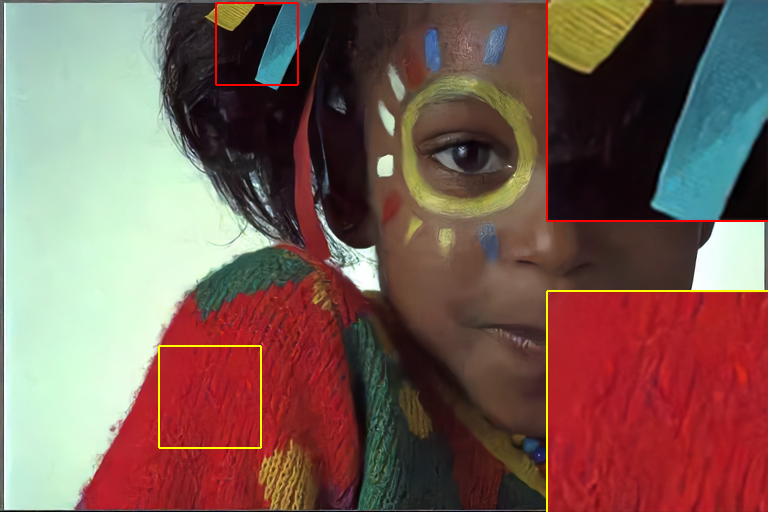} &
			\includegraphics[width=0.11\textwidth]{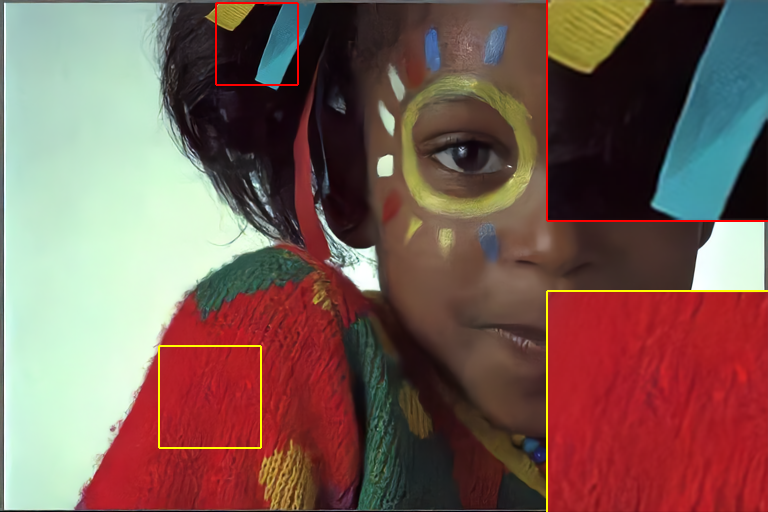} &
			\includegraphics[width=0.11\textwidth]{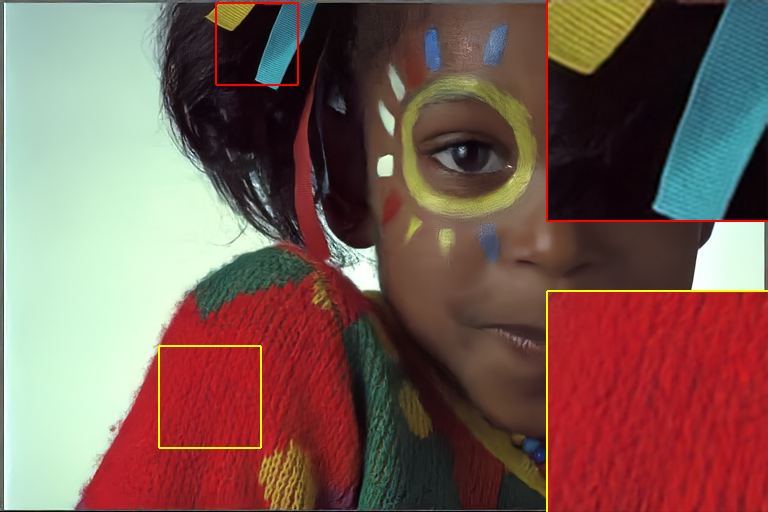} &
			\includegraphics[width=0.11\textwidth]{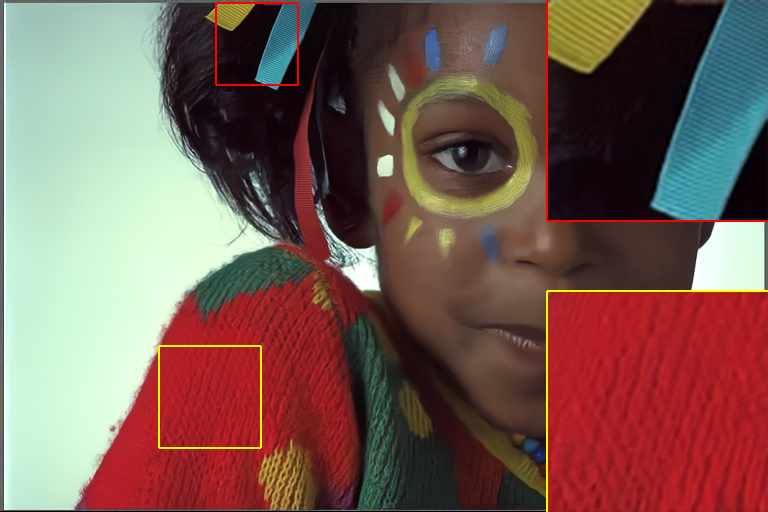}\\
		& (a) Ground truth & (b) Noisy ($\sigma=35$) & (c) BM3D-G~\cite{Davy2021BM3D} & (d) PMRID~\cite{Wang2020Practical} & (e) FDnCNN~\cite{Zhang2017DnCNN} & (f) FFDNet~\cite{Zhang2018FFDNet} & (g) Ours-FL (NLM) & (h) Ours-FL (BM3D) \\
		& CPSNR & 18.23 dB & 31.58 dB & 31.84 dB & 31.86 dB & 31.88 dB & 32.04 dB & \textbf{32.17} dB \\

        \multirow{1}{*}[18pt]{(C)} &
			\includegraphics[width=0.11\textwidth]{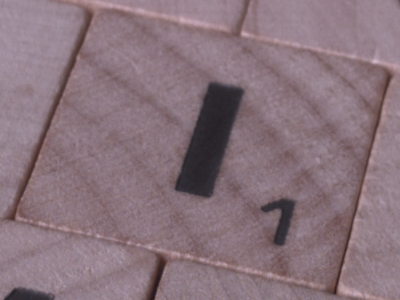} &
			\includegraphics[width=0.11\textwidth]{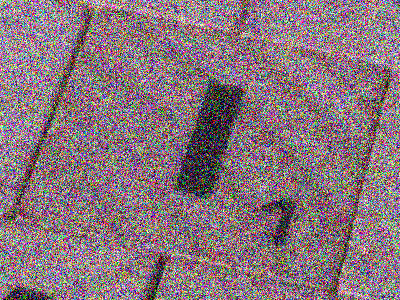} &
			\includegraphics[width=0.11\textwidth]{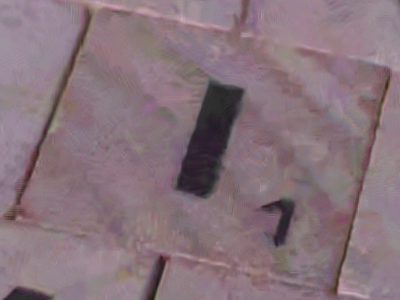} & 
			\includegraphics[width=0.11\textwidth]{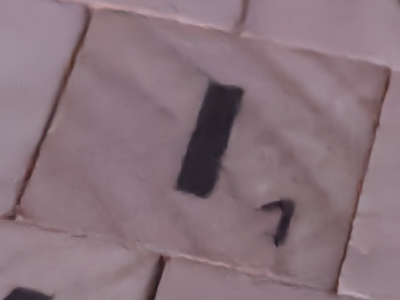} &
			\includegraphics[width=0.11\textwidth]{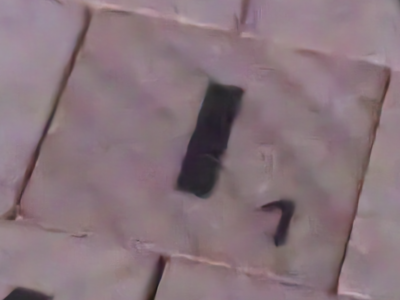} &
			\includegraphics[width=0.11\textwidth]{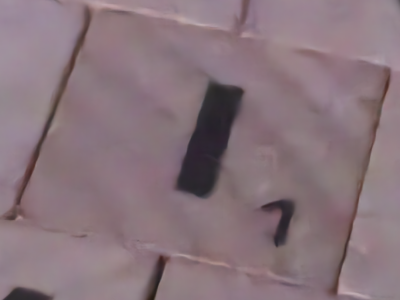} &
			\includegraphics[width=0.11\textwidth]{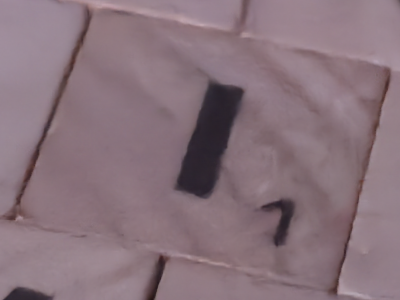} &
			\includegraphics[width=0.11\textwidth]{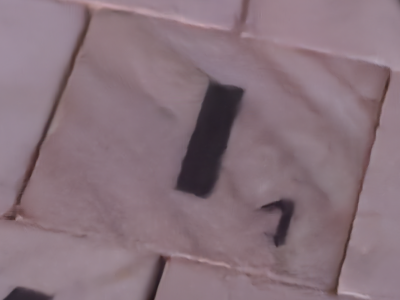}\\
			& (a) Ground truth & (b) Noise ($\sigma=75$) & (c) BM3D-G~\cite{Davy2021BM3D} & (d) PMRID~\cite{Wang2020Practical} & (e) FDnCNN~\cite{Zhang2017DnCNN} & (f) FFDNet~\cite{Zhang2018FFDNet} & (g) Ours-FL (NLM) & (h) Ours-FL (BM3D) \\
			& CPSNR & 11.66 dB & 33.17 dB & 36.23 dB & 34.71  dB & 35.23 dB & \textbf{36.29} dB & 36.25 dB
		\end{tabular}
		}
		
		\caption{Barbara, image 15 of \textit{Kodak} ($768\times512$) and image 156 of \textit{SIDD} ($4048\times3044$)  processed by different methods.}
		\label{SIDD}
	\end{figure*}

	\begin{table*}[t]
		\caption{Comparison of network complexity.}
		\label{complexity}
		\begin{center}
		\addtolength{\tabcolsep}{-1.2pt}
		\begin{tabular}{clcccccccccc}
			\hline
			 Image size & & NLM & BM3D & DnCNN & FDnCNN & DnCNN & FFDNet & PMRID & ADNet & Ours & Ours-FL  \\ 
			 & & -G~\cite{Davy2021BM3D} & -G~\cite{Davy2021BM3D} & \cite{Zhang2017DnCNN} & \cite{Zhang2017DnCNN} & $K = 10$ & \cite{Zhang2018FFDNet} & \cite{Wang2020Practical} & \cite{TIAN2020ADNet} & $K = 10$ & NLM/BM3D \\ \hline

			\multirow{4}{*}{\textit{$512\times512$}} & GFLOPs & -- & -- & 175.78 & 175.32 & 78.48 & 55.97 & 4.71 & 136.71 & 78.82 & 4.52  \\
			& Memory (MB)  & 65 & 71   & 904   & 841 & 773 & 747 & 785 & 905 & 839 & 802   \\
			& Time-GPU (s) & 0.03 & 0.09 & 0.17 & 0.15 & 0.07 & 0.05 & 0.03 & 0.13 & 0.07(+0.09) & 0.03(+0.03/0.09)  \\
			& Time-CPU (s) & -- & -- & 42.58 & 41.31 & 18.55 & 3.43 & 1.86 & 27.65 & 18.57(+0.09) & 1.82(+0.03/0.09)  \\ \hline
			\multirow{4}{*}{\textit{$1600\times1200$}} & GFLOPs & -- & -- & 1287.42 & 1284.10 & 574.83 & 409.93 & 34.47 & 1001.27 & 577.30 & 33.12  \\
			& Memory (MB)  & 137  & 184  & 3015 & 2579 & 2069 & 1421 & 1823 & 3011 & 2605 & 1867  \\
			& Time-GPU (s) & 0.16 & 0.66 & 1.21 & 1.05 & 0.50 & 0.35 & 0.21 & 1.13 & 0.50(+0.66) & 0.21(+0.16/0.66)  \\
			& Time-CPU (s) & -- & -- & 87.02 & 82.98 & 39.20 & 21.40 & 12.44 & 73.03 & 39.29(+0.66) & 9.19(+0.16/0.66)  \\ \hline
			\multicolumn{2}{c}{Model size (MB)} & 0.13 & 0.15 & 2.67  & 2.64 & 1.20 & 3.38 & 4.11 & 2.09 & 1.19(+0.15) & 3.87(+0.13/0.15)  \\ \hline
		
		\end{tabular}
		\end{center}	
	\end{table*}

	\subsection{Quantitative and Qualitative Comparison}
	We conducted comprehensive testing on five image sets in order to fully evaluate the performance of the algorithm. 
	Three flexible deep learning algorithms (flexible DnCNN (FDnCNN)~\cite{Zhang2017DnCNN}, FFDNet~\cite{Zhang2018FFDNet}, PMRID~\cite{Wang2020Practical}) were used for comparison. PMRID is a lightweight raw image  denoising algorithm for mobile devices, that we  retrained in order to compare it with our algorithm.

    Table \ref{Tab2} shows that our proposed method outperforms other flexible CNNs algorithms, especially for MIT moir\'{e} and Urban100, which are rich in details.	On the MIT moir\'{e} image set our proposed method beats other CNNs methods  by $0.75$dB. Nowadays, users and manufacturers prefer 4K resolution images. Therefore, we selected $25$ normal light images taken by Google Pixel ($4048\times3044$) from the SIDD dataset~\cite{Abdelhamed2018} for testing. 
    For these high resolution images, our proposed schemes maintain their advance.

	A comparison of Table~\ref{Tab1} and Table~\ref{Tab2} shows that the performance of FDnCNN has decreased compared with DnCNN. Similarly, Ours-FL (BM3D) has a slight decrease in CPSNR compared to Ours ($K=16$). 
	However, in the following evaluation of the cost of the algorithm, we find that these drops are worthwhile. 
	Indeed, it is more difficult to train with flexible parameters than fixed parameters, and the search for local optimal solutions consumes more resources.
	The comparison with \cite{Wang2020Practical} also shows that while direct application of lightweight networks can reduce the computational cost, its performance is lower in textured areas. Therefore, using BM3D for preprocessing can effectively reduce the difficulty and improve performance.

	Fig.~\ref{SIDD} compare the visual quality of the considered algorithms. 
	In Fig.~\ref{SIDD} (A), the texture  of the chair  is  messy  and the shadows on the pants are over-smoothed in the image restored by FDnCNN and FFDNet, while the texture  of the chair is perfectly preserved and the shadows are preserved by our method. The same conclusion can be drawn from Fig.~\ref{SIDD} (B) and (C).

From these results we conclude that our proposed method is good at preserving details and texture compared to CNNs algorithms. It also  creates fewer artifacts than non-local methods. In fact, the non-local part of our proposed framework can be replaced with any non-local denoising algorithm, as long as its processing speed can be guaranteed. 
Based on this, we also tested the use of NLM-G~\cite{Davy2021BM3D} as preprocessing algorithm. 
Compared to BM3D, NLM is computationally less expensive and therefore widely used in cameras. 
Ours-FL (NLM) yields significantly better results than NLM. On high resolution images, the NLM scheme is far less complex than the BM3D scheme (see Table \ref{complexity}).  Yet, its PSNR and visual quality is only slightly inferior to that of the BM3D scheme (see Table \ref{Tab2} and Fig.~\ref{SIDD} (C)).

\subsection{Network Complexity and Running Time}
	 The network complexity and running time were compared on a PC with Intel Core i7-9750H 2.60GHz, 16GB memory, and Nvidia GTX-1650 GPU. We tested images of two sizes: $512\times512$ (the most commonly used benchmark) and $1600\times1200$ (two megapixels, the maximum size that DnCNN can run on a 4G video memory device). 
	Four indicators: Floating point operations (FLOPs), Memory (MB), Time-GPU (s) and Time-CPU (s) are used for comparison.

	 The experimental results are shown in Table~\ref{complexity}. 
	 In terms of computation, our model is only $8.08\%$ of FFDNet and $2.58\%$ of FDnCNN. 
	 The significant reduction in the amount of calculation makes it possible to deploy 
	 on mobile terminals. 
	 In terms of run time, our algorithm retains the advantages on both CPU and GPU. As the resolution increases, the time of BM3D begins to increase, but NLM+lightweight CNN provides a valuable alternative, with an excellent trade-off between computational cost and visual quality.

	  For a higher cost solution with optimal image quality, BM3D remains the best choice. Since BM3D has  already been successfully implemented in several smart phones, a BM3D+lightweight CNN solution  is feasible.

	\section{Conclusion}
    In this letter, we explored a way to combine the advantages of  non-local denoising with those of  denoising CNNs. We saw that	non-locality recovers image texture, and that combining a nonlocal algorithm with a lightweight CNN corrects the image artifacts. The syncretic algorithm requires much less computational power than current CNNs and can therefore be used in mobile terminals.

	\newpage
    \balance
	\bibliographystyle{IEEEtran}
	\bibliography{ref.bib}
\end{document}